# On the nature of the melting line of *bcc* sodium

Eduardo R. Hernández* and Jorge Íñiguez

*Institut de Ciència de Materials de Barcelona (ICMAB–CSIC), Campus de Bellaterra, 08193 Barcelona, Spain*

Recent experiments have obtained the melting line of sodium up to pressures of about 130 GPa, finding that the melting line from the *bcc* phase reaches a maximum at a temperature of *c.a.* 1000 K and a pressure of 31 GPa, and at higher pressures the fusion temperature decreases continuously up to 118 GPa. Here we report results of a study based on first principles molecular dynamics, clarifying the nature of the maximum and subsequent decreasing behavior found in the melting line of sodium.

There has been considerable interest in the thermal behavior of the alkali metals Li and Na in recent years, particularly in what concerns their high pressure phases. At ambient pressure these elements have simple structures, and behave like free electron metals. However, it was first suggested by first principles calculations [1, 2], and then confirmed by experimental studies [3], that the phase behavior of Li and Na at high pressures leads to rather complex crystalline phases. It now seems that there were still further surprises: Gregoryanz *et al.* [4] recently measured the melting curve of Na up to pressures of 130 GPa, and discovered that around 31 GPa the melting line displays a maximum (at a temperature of around 1000 K); at higher pressures the melting line has a negative slope, which continues down to the pressure domain of the *fcc* phase (65 GPa$\leq P \leq$ 103 GPa), and only becomes positive again around 120 GPa, in the pressure range where solid Na is reported to have a complex cubic structure with 16 atoms in the unit cell, by which time the melting temperature falls down to 315 K, well below its value at ambient pressure (371 K).

Melting lines with negative slope are nothing new. Indeed, water has a decreasing melting line around ambient pressure, and the melting line of Si from the diamond phase also presents a negative slope. But the range of pressures in which the melting line of water decreases is small (around 0.2 GPa), and in Si it is around 10 GPa [5]. According to the measurements of Gregoryanz *et al.* [4] the melting line of Na decreases in a range of pressures of nearly 90 GPa, from 31 GPa (where the maximum is located) up to 118 GPa, with a possible flat region (not resolved by the experiments) around the location of the *bcc*-*fcc*-liquid triple point.

In this work we address the thermal properties of Na from first principles electronic structure calculations. Our aim is to clarify the reasons that lead to the occurrence of the maximum along the melting line of *bcc* Na found at 31 GPa.

There are two possible scenarios that could account for the presence of the experimentally observed maximum: *i)* a phase transition occurs in one of the two phases separated by the melting line; *ii)* there are no phase transitions on either side of the melting line close to the maximum; rather the two phases have different compressibilities, and hence react differently to the external pressure, reaching equal densities (and thus molar volumes) at the maximum of the melting line. Let us first consider scenario *i)*. If some phase transition occurs, it must occur in the liquid, as in the pressure range where the maximum of the melting line is found, the solid is reported to be in the *bcc* structure. Besides, if there was a phase transition in the solid, one would expect that it is a transition to a phase having a lower molar volume, since the imposed pressure is high, and this would result in the slope of the melting line becoming more positive, rather than negative. If indeed a phase transition takes place in the liquid, and it is a first order phase transition, we would expect a discontinuity in the pressure derivative of the melting line at the maximum. The pressure derivative of the melting line is given by the Clausius-Clapeyron equation, which reads:

$$\frac{dT_m}{dP} = T_m \frac{\Delta V}{\Delta H}, \tag{1}$$

where $T_m$ is the melting temperature, $P$ the pressure, $\Delta V = V_l - V_s$ is the difference of molar volumes and $\Delta H = H_l - H_s$ the difference of molar entropies. Both $V_l$ and $H_l$ would change abruptly if there is a first order phase transition in the liquid at the maximum of the melting line, and hence a discontinuous change in the slope of the melting line would result. In this case, the maximum of the melting line would in fact be a triple point, where the *bcc* solid phase would coexist with two different liquid phases. As pointed out by Gregoryanz *et al.* [4], this is actually the situation encountered in P [6, 7], where a first order liquid-liquid ($\ell - \ell$) transition occurs between a molecular and a polymeric liquid, which causes a discontinuous change in the slope of the melting line of the orthorhombic solid phase.

According to the second possible scenario, to the left of the maximum in the melting line we would have the common situation of a solid denser than the liquid into which it melts, also being less compressible. As the pressure is increased the liquid phase reduces its molar volume more rapidly than the solid, as it is more compressible, and hence, by virtue of Eq. (1), the slope of the melting line is gradually reduced, until it becomes zero, when the molar volumes of the liquid and solid phases become equal. In this case the pressure derivative would not present a discontinuity at the maximum.

The experimental data of Gregoryanz *et al.* [4] does not allow to discriminate between scenarios *i)* and *ii)* above, because the melting temperature was not measured at a sufficiently small pressure interval to allow a clear distinction. Nevertheless, Gregoryanz and coworkers suggested the existence of a phase transition leading to different structural and electronic properties in the liquid on either side of the melting curve maximum, with a non-monotonic change in the number of nearest neighbors and interatomic distances. In this Letter we analyze this question from the perspective of first principles molecular dynamics simulations.

Our calculations have been performed with cells containing 64 atoms of Na, both in the case of the liquid and solid *bcc* phases, considering a series of different volumes and temperatures. We used the Generalized Gradient Approximation [8] to Density Functional Theory as implmented in the VASP (Vienna Ab initio Simulation Package) [9] code. We employed the Projector Augmented Wave (PAW) method of Blöchl [10] as implemented in VASP [11]. The $2p$ and $3s$ electrons were included in the valence; at the large pressures considered in this study, it is necessary to incorporate the *2p* electrons in the valence to avoid an artificial dimerization of Na in the liquid phase. We used a basis set of plane waves up to a 260 eV cutoff, which we checked renders sufficiently converged results, and a grid of $2 \times 2 \times 2$ **k**-points to sample the Brillouin zone, generated according to the Monkhorst-Pack [12] procedure. We have systematically evaluated the effect of using a denser $(4\times4\times4)$ **k**-point grid and the inclusion of the *2s* electrons in the valence, by performing calculations on a subset of the structures resulting from our simulations. These effects are small, and amount to a rigid shift in energies and pressures, which we have included in all the results discussed below.

In our MD simulations we have used a time step of 1 fs, which is sufficiently short to provide a stable numerical integration of the equations of motion. First we equilibrated the system during a time of 1 ps, in which the velocities were scaled at every time step in order to drive the system towards the desired target temperature. The equilibration run was then followed by the production run, which was 9 ps long, and during which the dynamics of the atoms, coupled to a Nosé-Hoover [13, 14] thermostat, were monitored. This allowed us to simulate each system in canonical, or constant NVT, conditions.

We have carried out simulations of the solid and liquid phases in varying conditions of volume and temperature. Firstly, we have simulated both phases along the T=1000 K isotherm, at volumes corresponding to pressures that bracket the experimentally observed maximum, at around 31 GPa. At this temperature, the solid phase should be metastable, except possibly at the pressure of the melting curve maximum, where the melting temperature is approximately 1000 K. We also performed simulations at conditions of temperature and pressure which approximately mimic those of the experimental melting curve. We stress that in this study we did not attempt to calculate the theoretical melting curve. In Fig. (1) we have plotted the difference of the average pressure of the liquid ($P_l$) and the solid ($P_s$) phases as a function of the volume; also plotted in the inset is the variation of the pressure with the volume along the T=1000 K isotherm for the liquid phase. As can be seen in the plot, when the pressure corresponds to a value to the left of the melting curve maximum, we observe that the pressure of the liquid phase is higher than that of the solid phase. This occurs both at the temperatures corresponding to the experimental melting curve, and when the temperature is fixed at 1000 K. Since both phases were simulated at the same volume, this implies that in conditions of constant pressure the liquid phase would have a larger volume, which is consistent with the positive derivative of the melting curve in this region. On the other hand, when the volume is fixed at values corresponding to pressures to the right of the melting curve maximum, the situation is reversed, i.e. the solid phase has a larger pressure ($P_l - P_s$ is negative). This indicates that at equal pressures, this phase would have the larger volume, which is again consistent with the experimental finding of a negative slope in the melting curve at pressures above 31 GPa. In the conditions which resemble more closely those of the melting curve maximum, we observe that the pressure of both phases is very nearly the same, with a difference of less than 1 kB between them. Hence, we can conclude that the results of our simulations are compatible with the experimental observations [4].

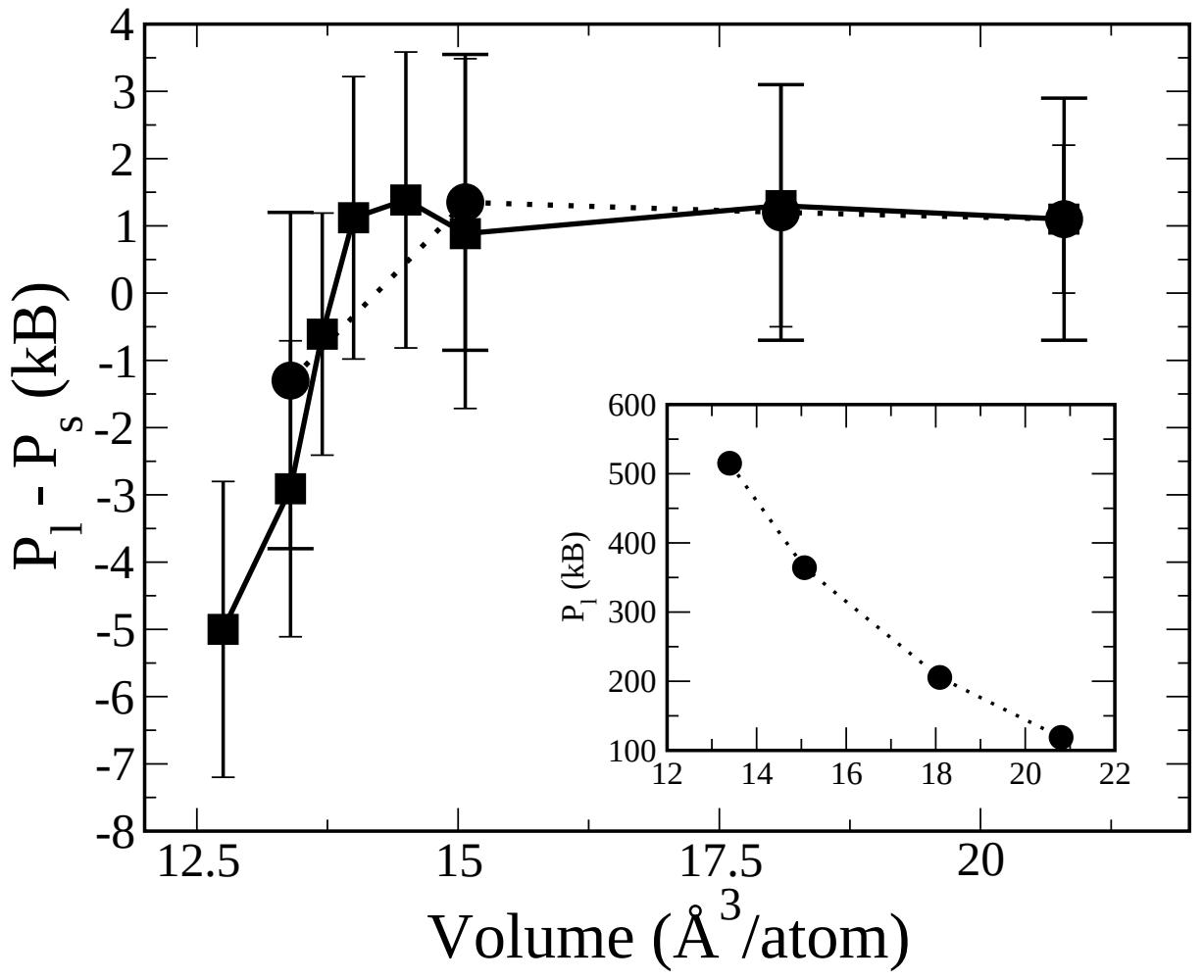

FIG. 1: Pressure difference between the liquid ($P_l$) and solid ($P_s$) phases as a function of volume. The continuous line with square symbols was obtained at conditions which approximately follow the experimental melting line, while the dotted line with circles was obtained along the T=1000 K isotherm. The error bars reflect the standard deviation of the average pressure in our simulations. The inset shows the pressure of the liquid phase in the same range of volumes, along the T=1000 K isotherm; the pressure of the solid phase is indistinguishable from that of the liquid on the scale of the plot.

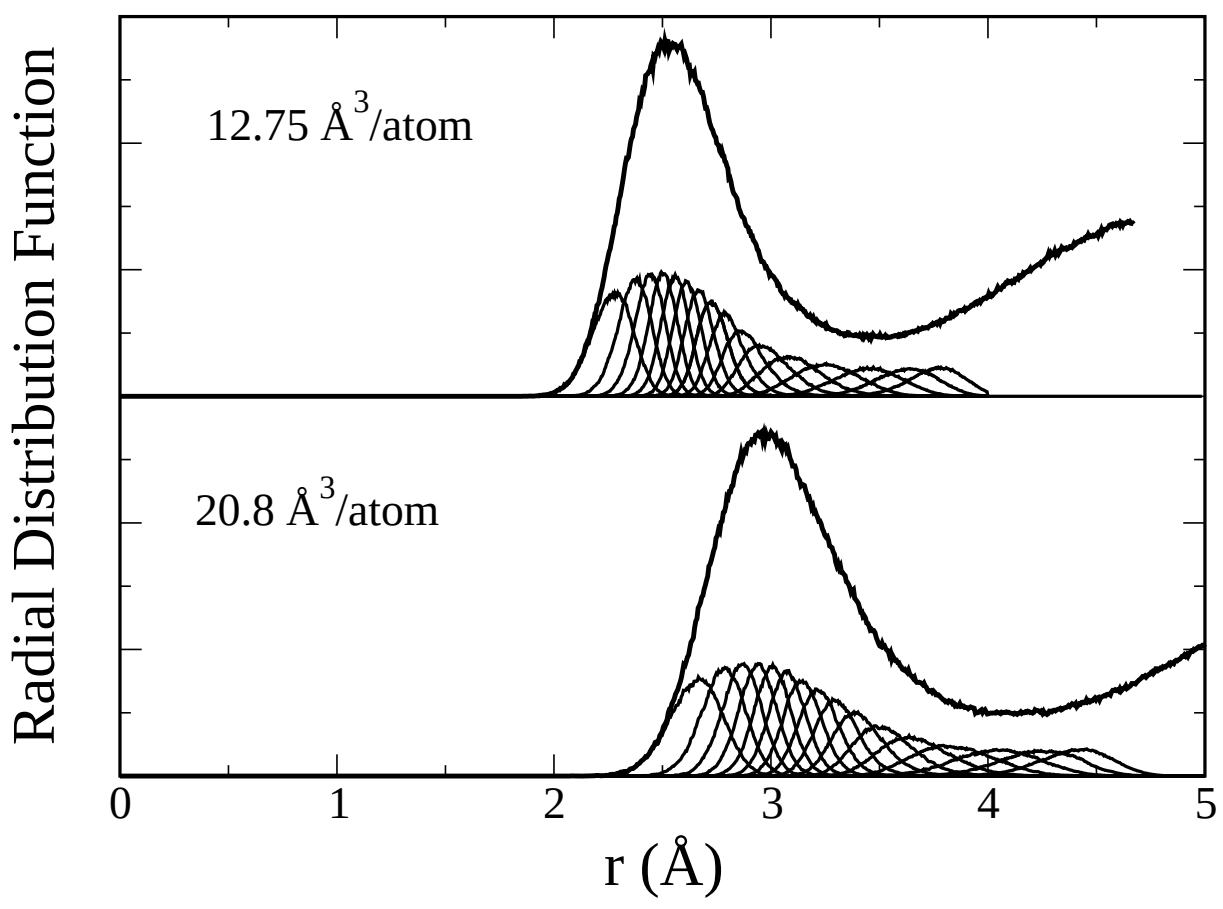

FIG. 2: Radial distribution function and pair distribution functions of the sixteen nearest neighbors in the liquid phase. Results obtained at volumes of 20.8 Å$^3$/atom and at 12.75 Å$^3$/atom are shown.

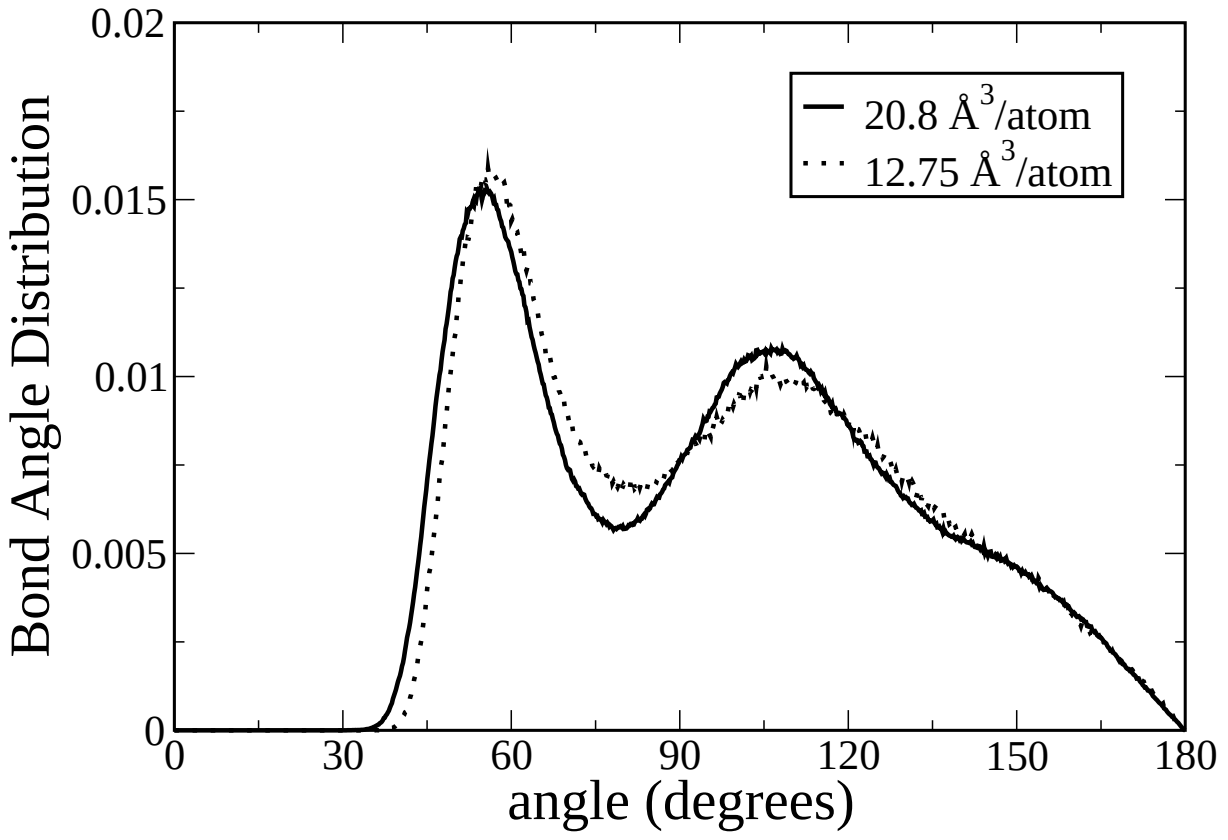

FIG. 3: Bond-angle distribution function of the liquid at volumes of 20.8 Å$^3$/atom and 12.75 Å$^3$/atom.

The abruptness of the change in the pressure difference, $P_l - P_s$, made evident in Fig. (1) would seem to imply that some drastic change takes place in the liquid structure as the volume is reduced. We have carefully analyzed the structure of the liquid phase in search of any signatures of change; for example, in Fig. (2) we plot the radial distribution function (RDF) and the pair distribution functions (PDF) of the sixteen nearest neighbors for the liquid phase at volumes of 20.8 Å$^3$/atom and 12.75 Å$^3$/atom, corresponding to the rightmost and leftmost points in Fig. (1), respectively. At the higher pressure (lower volume) the system is naturally more compressed; the first peak of the RDF shifts to shorter distances by about 0.5 Å. The PDFs of the sixteen nearest neighbors also shift accordingly, but, as is evident from the plot, the compression causes no change in the average coordination of the liquid. Indeed, in both the low and high pressure liquid we can see that the maximum of the PDF corresponding to the fourteenth nearest neighbor is very close to the position of the first minimum in the RDF, so we can say that on average each atom has 13-14 neighbors in the liquid, both at the lower and higher pressures. We have also analyzed the bond-angle distribution functions in these two extreme cases, and they are illustrated in Fig. (3). It can be seen there that at the higher pressure the lower part of the distribution shifts towards slightly larger angles, but there is otherwise no major change in the distribution. A detailed comparison with the angle distribution corresponding to ideal *bcc* and *fcc* phases revealed no clear short range order in the simulated liquids.

Given that the calculated RDFs and angle distributions are not very informative regarding the local structure of the liquid, we resorted to other analysis strategies proposed in the literature, that are more sensitive to the local environment. Specifically, we have employed the rotational invariant parameters of Steinhardt *et al.* [15], and those of Rodriguez de la Fuente and Soler [16]. We have also tried the criteria recently proposed by Ackland and Jones [17], but none of these procedures seems to make evident any noticeable structural change in the liquid as the pressure is increased. In Fig. (4) we have plotted the frequency histograms of the $\overline{W}_6$ rotationally invariant parameter defined by Steinhardt *et al.* [15], which is particularly sensitive to the type of environment, calculated for clusters formed by an atom and its nearest neighbors, obtained from simulations at volumes of 20.8 Å$^3$/atom and 12.75 Å$^3$/atom. The two histograms are seen to overlap almost perfectly, which we interpret as an indication that there is no change in the local environment of the liquid at these different densities. We have also indicated in Fig. (4) the values of the $\overline{W}_6$ parameter for the perfect icosahedral, *fcc* and *bcc* clusters. It is seen that the perfect icosahedral cluster is practically never observed in our simulations at either density, although clusters with values of $\overline{W}_6$ below -0.05 can be said to have some icosahedral character [18]. The histograms peak close to the value expected for the *fcc* cluster; note however that the ratio of frequencies at the values expected for the perfect *fcc* and *bcc* clusters do not change. The perfect *hcp* cluster would give a value for $\overline{W}_6$ very close to that of the *fcc* cluster. The main conclusion from this analysis is that the local atomic environment of the liquid does not change as the volume is reduced from 20.8 Å$^3$/atom to 12.75 Å$^3$/atom along the melting line.

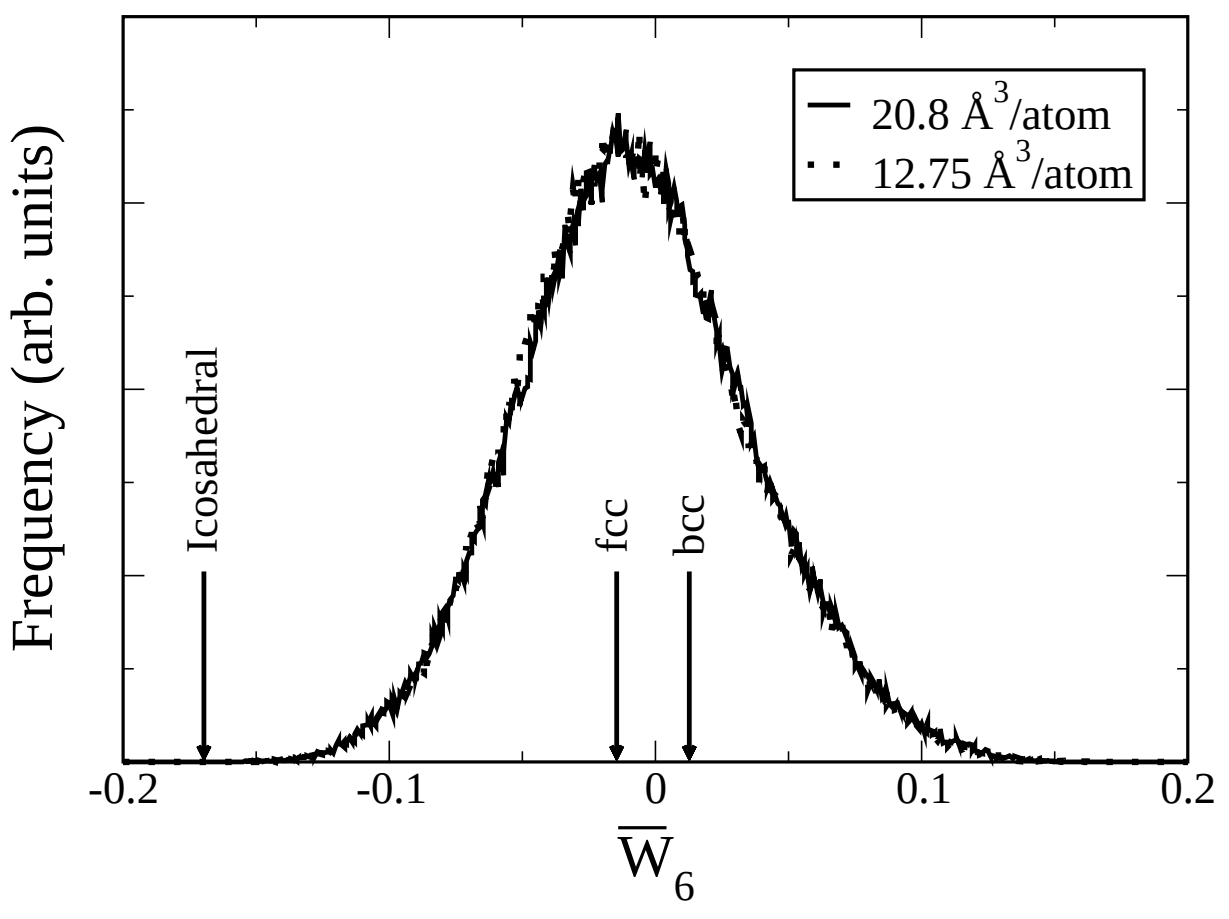

FIG. 4: Frequency histogram of the $\overline{W}_6$ parameter of Steinhardt *et al.* [15], for the clusters found in l-Na at densities of 20.8 Å$^3$/atom and 12.75 Å$^3$/atom.

We have also analyzed the electronic structure of the liquid at different pressures, but again failed to observe any marked difference, other than a shift towards higher energies as the pressure is increased, and in any case nothing indicative of a change in the pattern of bonding of the liquid.

In view of the fact that no abrupt structural transition becomes apparent in the liquid phase as the pressure is increased along the T=1000 K isotherm, nor along the melting line, it seems that the maximum in the melting curve of the *bcc* phase observed experimentally [4] must be due to the second kind of scenario mentioned earlier (scenario *ii)*), namely, that the liquid is more compressible than the solid, and therefore reacts to the imposed external pressure by reducing its volume more rapidly that the solid phase, until eventually the volumes of solid and liquid become identical at the maximum of the melting curve. Beyond this point, the molar volume of the liquid phase becomes smaller than that of the solid, and hence, by virtue of Eq. (1), the melting line develops a negative slope. This situation persists in the range of stability of the *bcc* phase (up to about 65 GPa). It is remarkable, however, that even beyond the *bcc*-*fcc* transition, which is accompanied by a reduction of the molar volume in the solid, the negative slope of the melting line is maintained. This implies that the compressibility of the liquid phase remains lower than that of the *fcc* phase, although close to the right of the *bcc*-*fcc*-liquid triple point the melting line appears to be rather flat. It is even conceivable that in this region the melting line briefly recovers a positive slope and reaches a second local maximum, before decreasing again. In fact, this situation seems to be actually encountered in the melting line of Cs [19], where two maxima appear, one on each side of the *bcc*-*fcc*-liquid triple point. Further experimental or theoretical studies would be required to resolve this intriguing possibility.

Note that, at variance with what has been hypothesized in the literature [4], our results indicate that the melting-line maximun of Na is markedly different in nature from that of P, as the latter is the result of a $\ell - \ell$ transition that does not occur in the Na case, at least in the range of pressures we have explored. In retrospect, such a difference may not be so surprising: the $\ell - \ell$ transition found in P is closely related, from the point of view of the system's short range order, to a solid-solid transition that takes place at *slightly* higher pressures (i.e. about 5 GPa higher). In the case of Na, the hypothetical $\ell - \ell$ transition would occur at about 31 GPa, while the closest, presumably associated, solid-solid transformation (i.e. *bcc* to *fcc*) takes place at about 65 GPa. It would have been very striking to find that the liquid acquires at 31 GPa the short range order that becomes stable in the solid phase only at pressures as high as 65 GPa. Comparison with the case of Cs is also enlightening. As pointed out above, the melting line of Cs shows two local maxima [19] on either side of the Cs I(*bcc*)–Cs II(*fcc*)–liquid-Cs triple point, occuring at 2.2 GPa, yet no $\ell - \ell$ transition has been reported [20] in this range of pressures. A $\ell - \ell$ does indeed occur at 3.9 GPa [20], but this is clearly linked to the Cs II–Cs III transition, and does not account for the local maxima close to the Cs I–Cs II–$\ell$-Cs triple point.

To summarize, we have conducted FPMD simulations of Na in the liquid and solid *bcc* phases at an extended range of temperature and volume conditions. Our results are consistent with the experimental observation of a melting-line maximum at about 30 GPa [4]. We find that the liquid phase is more compressible than the solid, implying that the chemical potential of the liquid is increased with pressure more slowly than that of the *bcc* phase. The molar volumes of both phases become equal in the range of 30-40 GPa, resulting in a maximum of the melting line, and at higher pressures the molar volume of the liquid phase falls below that of the solid. We do not find, however, any indication of a structural or electronic phase transition in the liquid phase in the range of pressures considered, and therefore conclude that the re-entrant behavior observed in this case is solely attributable to the different compressibilities of the solid and liquid phases. Our results serve also as a reminder that the presence of maxima along the melting line do not necessarily imply the presence of a liquid-liquid phase transition.

**Acknowledgments**

We acknowledge financial support by the Spanish Ministry of Science and Education through project BFM2003-03372-C03-03 and through a Ramón y Cajal Fellowship (JI); and from the Catalan Regional Government through project 2005SGR683. Simulations were performed at the Mare Nostrum computer, Barcelona Supercomputer Centre (BSC).

---

* email: ehe@icmab.es